\documentclass{emulateapj}

\usepackage{amssymb, amsmath, apjfonts, natbib, color}

\def\degree{{\circ}}
\newdimen\digitwidth
\setbox1=\hbox{0}
\digitwidth=\wd1
\catcode`"=\active
\def"{\kern\digitwidth}

\slugcomment{Accepted by ApJ Letter}

\begin{document}

\title{MAPPING THE THREE-DIMENSIONAL ``X-SHAPED STRUCTURE'' IN MODELS OF THE GALACTIC BULGE}

\author{Zhao-Yu Li\altaffilmark{1} and Juntai Shen\altaffilmark{1, 2}}

\altaffiltext{1}{Key Laboratory for Research in Galaxies and Cosmology, Shanghai Astronomical Observatory, Chinese Academy of Science, 80 Nandan Road, Shanghai 200030, China}
\altaffiltext{2}{Correspondence should be addressed to Juntai Shen: jshen@shao.ac.cn}

\begin{abstract}

Numerical simulations have shown that the X-shaped structure in the Milky Way bulge can naturally arise from the bar instability and buckling instability. To understand the influence of the buckling amplitude on the morphology of the X-shape, we analyze three self-consistent numerical simulations of barred galaxies with different buckling amplitudes (strong, intermediate and weak). We derive the three-dimensional density with an adaptive kernel smoothing technique. The face-on iso-density surfaces are all elliptical, while in the edge-on view, the morphology of buckled bars transitions with increasing radius, from a central boxy core to a peanut bulge and then to an extended thin bar. Based on these iso-density surfaces at different density levels, we find no clear evidence for a well-defined structure shaped like a letter X. The X-shaped structure is more peanut-like, whose visual perception is probably enhanced by the pinched inner concave iso-density contours. The peanut bulge can reproduce qualitatively the observed bimodal distributions which were used as evidence for the discovery of the X-shape. The central boxy core is shaped like an oblong tablet, extending to $\sim$ 500 pc above and below the Galactic plane ($|b| \sim 4^\degree$). From the solar perspective, lines of sight passing through the central boxy core do not show bimodal distributions. This generally agrees with the observations that the double peaks merge at $|b| \sim 4^\degree - 5^\degree$ from the Galactic plane, indicating the presence of a possibly similar structure in the Galactic bulge.

\end{abstract}


\section{INTRODUCTION}
\label{sec:intro}

Commonly found in disk galaxies, bars play an important role in redistributing energy and angular momentum, thus shaping the galaxies through secular evolution \citep{kor_ken_04, sellwo_14}. In edge-on disk galaxies, the boxy/peanut-shaped (B/PS) bulges are frequently observed and suggested as bars viewed edge-on; their relatively flat light distribution, cylindrical rotation, and sometimes X-shaped morphology are similar to bars rather than classical bulges \citep{kui_mer_95, bur_fre_99}. Numerical simulations confirm that, a bar, once formed as a result of the bar instability, quickly buckles in the vertical direction to form a B/PS bulge in the edge-on view \citep{com_san_81, combes_etal_90, raha_etal_91, martin_etal_06}. This buckling instability is very efficient at heating the disk vertically to generate peanut shaped bulges \citep{debatt_etal_06}. However, the influence of buckling instability on bar morphology is still not well understood.

The boxy part of an $N$-body bar is shorter than the bar as a whole \citep{kor_ken_04, athana_05, debatt_etal_05}. Thus, a thickened bar seems to be composed of two components, i.e., the inner thick B/PS bulge at half bar length and an extended thin bar. In moderately inclined galaxies, evidences of the B/PS bulges may be identified based on some kinematic and isophotal imprints \citep{debatt_etal_05, erw_deb_13}. Close to the face-on view, it is difficult to distinguish the B/PS bulge from the bar.

In the Milky Way, both star counts and gas/stellar kinematics confirm the existence of a bar in the bulge region, with its detailed properties still under active debate \citep{bli_spe_91, bis_ger_02, ratten_etal_07, shen_etal_10, cao_etal_13, pietru_etal_14}. Aiming to understand the kinematics and the photometry of the Galactic bulge, \cite{shen_etal_10} developed a simple self-consistent numerical model, in which a bar forms spontaneously from a massive cold disk and buckles subsequently in the vertical direction, resulting in a B/PS bulge. From the solar perspective, this model matches the Bulge Radial Velocity Assay (BRAVA) kinematics and Cosmic Background Explorer (COBE) near-infrared photometry of the Galactic bulge strikingly well. Such a simple model is highly successful in matching many aspects of the Milky Way bulge (see the recent review by \citealt{she_li_15}). A significant classical bulge is also ruled out. From the galaxy formation point of view, the Milky Way is probably a pure disk galaxy mainly shaped via internal secular evolution.

The discovery of the split of red clump stars (RC) \citep{mcw_zoc_10, nataf_etal_10, saito_etal_11} is interpreted as evidence for a vertical X-shaped structure \citep{mcw_zoc_10}. This X-shaped structure seems to exist at $|b| \ge 4^\degree$, and may be off-centered \citep{portai_etal_15a}. As shown in \cite{li_she_12} (also see \citealt{ness_etal_12}), the observational signatures of the X-shape can be well reproduced with the B/PS bulge bar model in  \cite{shen_etal_10}.

Previous star count studies have inferred another planar long bar component, which seems offset from the bulge bar \citep[e.g.,][]{benjam_etal_05, lopezc_etal_07}. With a numerical barred model, \cite{mar_ger_11} have suggested that observations of this twisted long bar can be reproduced by the leading part at the bar ends. However, based on infrared photometry of the Galactic bulge stars, \cite{wegg_etal_15} inferred the existence of a planar long bar aligned with the Galactic B/PS bulge bar.

A comprehensive analysis of bar morphology under different buckling amplitudes can help to better understand the X-shape and B/PS bulge of both the Milky Way and external galaxies. Here we derive the three-dimensional iso-density of bars with different extents of buckling amplitudes. The simulation configurations are briefly described in Section~\ref{sec:model}, with the corresponding iso-density surfaces shown in Section~\ref{sec:results}. The results are discussed in Section~\ref{sec:discussion} and summarized in Section~\ref{sec:summary}.

\begin{figure}
\epsscale{0.9}
\plotone{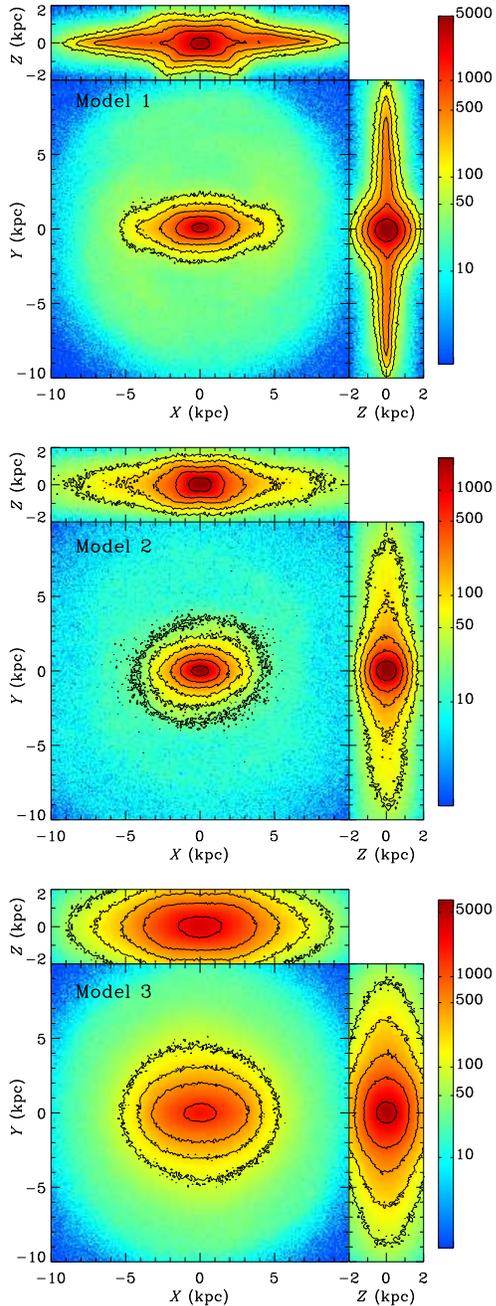}
\caption{Illustration of the three models used in our work. Color represents the projected particle number density per pixel, which is scaled logarithmically.}
\epsscale{1.}
\label{fig:show_model}
\end{figure}

\begin{figure}
\plotone{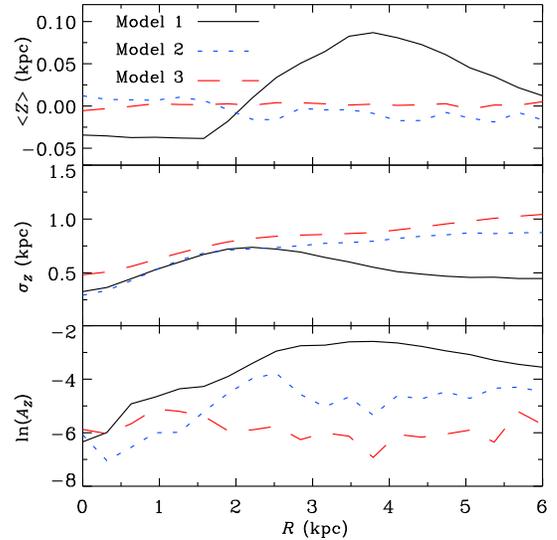}
\caption{Radial profiles of the average height $<Z>$ (top row), height dispersion $\sigma_Z$ (middle row), and buckling amplitude ln$(A_Z)$ (bottom row) of the three models.}
\label{fig:buckling_amp}
\end{figure}

\section{SIMULATIONS}
\label{sec:model}

In this work, we analyze three $N$-body simulations with different buckling strengths, which are shown in Figure~\ref{fig:show_model} with the buckling strength decreasing from the top to bottom panels. Initially, the three models are featureless exponential disks. In Model 1, two million disk particles evolve in a live dark matter halo, which consists of 2.5 million particles with the compressed King profile ($\Psi(0) / \sigma^2 = 3$ and $r_{\rm c} = 10R_{\rm d}$, see \citealt{sel_mcg_05} for details of adiabatic compression). Model 2 is a successful Galactic bulge model in \cite{shen_etal_10}, where one million disk particles rotate in a rigid dark matter halo potential. In both models, the bar quickly forms and sequentially buckles to induce an inner B/PS bulge; the B/PS bulge in Model 1 is more prominent than Model 2, because the buckling amplitude is stronger. Both models, especially Model 2, have been widely used in understanding the properties of the Galactic bulge and disk \citep[e.g.,][]{shen_etal_10, li_she_12, li_etal_14, molloy_etal_15a, molloy_etal_15b, nataf_etal_15, qin_etal_15}. 

We also include Model 3, a thick disk with a weakly buckled bar. The initial configuration is similar to the model R5 in \cite{debatt_etal_05} \citep[also see][]{gardne_etal_14}. Two million disk particles evolve in a rigid dark matter halo; the initial disk thickness is 6 times larger than that in Model 2. After formation, the bar barely buckles in the vertical direction. 

The three models have been properly scaled in terms of size ($R_{d,0}$) and velocity ($V_c$) to roughly match that of the Milky Way properties. From the solar perspective, both Models 1 and 2 demonstrate clear bimodal features in stellar distance distributions along different lines of sight toward the Galactic bulge as in \cite{mcw_zoc_10} and \cite{nataf_etal_10}. This bimodal feature is absent in Model 3.

In the top panel of Figure~\ref{fig:buckling_amp}, Model 1 has the largest average height $<Z>$ inside the bar region. For Model 2, the average height is negative, while for Model 3, it is close to 0. This is consistent with the result in \cite{sel_mer_94} that stronger buckling amplitudes tend to enhance asymmetries with respect to the disk plane. Outside the bar region, the outer disk thickness traced by $\sigma_Z$ (middle panel) increases from Model 1 to Model 3, which is expected from our initial disk configuration. In the bottom panel, the $m = 2$ bending amplitude within each radial bin is estimated as in \cite{sel_mer_94} and \cite{debatt_etal_06}:
\begin{eqnarray}
A_Z = \frac{1}{N}|\sum_{j}z_j e^{i2\phi_j}|.
\end{eqnarray}
According to the maximum $A_Z$ within the bar region, the buckling amplitude is the strongest in Model 1, intermediate in Model 2, and the weakest in Model 3. In particular, $A_Z$ in the bar region of Model 3 has remained small at all times, indicating that it experienced almost no buckling during the evolution of this model.

\section{ISO-DENSITY SURFACES OF THE BAR}
\label{sec:results}

To create the iso-density surfaces from numerical simulations, we smooth particles with a quadratic kernel function ($f(r) = 1 - (r / h)^2$), where $h$ is the kernel size. However, a uniform kernel for each particle is usually not suitable, because of the inhomogeneous distribution of particles and their limited numbers. With very small kernel sizes, the surfaces are overly noisy, while the very large kernel size tends to create over-smoothed surfaces. Therefore, in dense regions, smaller kernels should be applied to avoid over-smoothing, while in sparse regions, larger kernels are preferred. We adopt an adaptive kernel smoothing technique to optimize the individual kernel sizes \citep{silver_86, she_sel_04}. This technique starts with an initially uniform kernel size for each particle to roughly estimate the density in a pilot run. Then the individual kernel size is adjusted with a proper scaling factor for each particle; the kernel size shrinks in dense regions, and expands in sparse regions. Several different values for the initial kernel size are tested. Empirically, the initial kernel size of 1.0 kpc gives a good trade-off between noisiness and bias due to over-smoothness. The adaptive nature in our kernel smoothing technique ensures that the 3D density surface is not very sensitive to the exact initial values of the kernel sizes.

\begin{figure}
\plotone{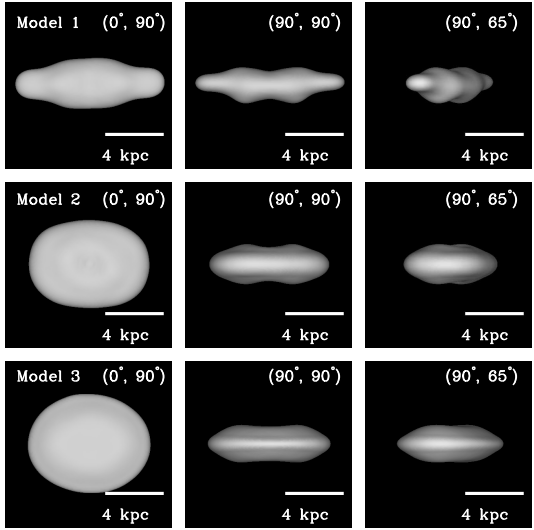}
\caption{Iso-density surfaces of Models~1, 2, and 3 are shown from top to bottom rows, respectively. For each model, from left to right, the iso-density surfaces are projected from different viewing angles in the format $(i, \Phi_{\rm bar})$; $i$ and $\Phi_{\rm bar}$ represent the disk inclination angle and the bar direction, respectively.}
\label{fig:iso-density_out}
\end{figure}

The iso-density surfaces of Models 1, 2, and 3 at large radial ranges are shown in the top, middle, and bottom rows of Figure~\ref{fig:iso-density_out}, respectively. The left, middle, and right columns correspond to face-on, side-on, and almost end-on viewing angles. The disk inclination angle ($i$) and the angle of the bar major axis with respect to the line of sight ($\Phi_{\rm bar}$) are labeled in each panel in the format ($i, \Phi_{\rm bar}$). The buckled bar is composed of an inner peanut bulge and an extended thin bar (e.g., \citealt{athana_05}); this feature is not clear in the weakly buckled Model 3. From the top (Model 1) and middle (Model 2) rows, the peanut bulge roughly extends to $0.5 R_{\rm bar}$, where $R_{\rm bar}$ is the radial size of the bar. In the face-on view, the ellipticity of the extended thin bar is slightly higher than the peanut bulge. If included in photometric decompositions, this component may potentially influence the best-fit bulge's mass/luminosity \citep[e.g.,][]{laurik_etal_14}.

\begin{figure}
\plotone{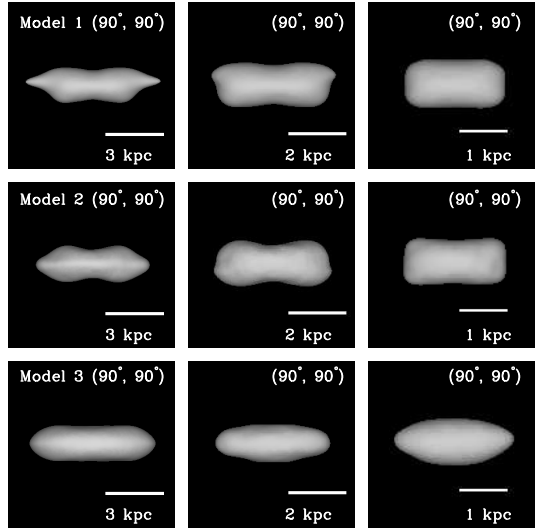}
\caption{Iso-density surfaces in increasingly higher density regions of the three models. For each model, from left to right, the side-on views $(i = 90^\degree, \Phi_{\rm bar} = 90^\degree)$ of the iso-density surfaces at 3, 2, and 1 kpc are illustrated, respectively.}
\label{fig:iso-density_in}
\end{figure}

Figure~\ref{fig:iso-density_in} shows the side-on view of the iso-density surfaces ($i = 90^\degree, \Phi_{\rm bar} = 90^\degree$) in increasingly higher density regions from 3 kpc (left column) to 1 kpc (right column) for the three models. Apparently, the extended thin bar disappears at the highest density level. The iso-density surfaces of Models 1 and 2 are peanut-like at $\sim$ 3 kpc. Iso-density surfaces become concave towards the central region. From face-on projection, their morphologies are all elliptical with a similar ellipticity. At $\sim$ 1 kpc shown in the right column, the peanut bulge transitions to a side-on boxy core, whose shape is like an oblong tablet. The iso-density surfaces of Model 3 in the edge-on view remain similar to ellipses without the presence of a peanut bulge or a boxy core. Therefore, the iso-density surfaces reveal more morphological structures in the buckled bar than were previously thought to exist.

After the buckling instability, the bar structure consists of a newly identified central boxy core, a peanut bulge, and an extended thin bar. These terminologies mainly come from their side-on appearances. The thin bar extends to the ends of the overall bar structure (half length $\sim$ 4 kpc) with vertical thickness similar to the outer disk. The peanut bulge resides in the inner half of the bar (extending to $\sim$ 2 kpc), which is almost twice the thickness of the extended thin bar. The half length of the boxy core is $\sim$ 1 kpc, which extends to $\sim$ 500 pc above the disk plane. Depending on the buckling strength, the relative significance of each of the three components are also different. The extended thin bar and the peanut component are clearly seen in Model 1, which become slightly weaker in Model 2, and almost absent in Model 3. In the Milky Way, the long bar identified in \cite{wegg_etal_15} seems twice the size of the X-shaped/peanut bulge, which agrees quite well with expectations from numerical models.

As in \cite{wegg_etal_15}, we derive the number density profiles as a function of Galactic latitude ($b$) in the bar region. Compared to the observed scale height in \cite{wegg_etal_15}, both Models 1 and 2 produce consistent results in the inner region of the bar ($-10^\degree < l < 10^\degree$); the scale height varies from $\sim 2.5^\degree$ (at $l = 0^\degree$) to $\sim 5.5^\degree$ (at $l = 10^\degree$), corresponding to $\sim 350$ pc and $550$ pc, respectively. While in the outer region of the bar ($10^\degree < l < 30^\degree$), the typical scale heights are $\sim 4^\degree$ (400 pc) for Model 1 and $\sim 8^\degree$ (600 pc) for Model 2, which are larger than the result ($\sim$ 200 pc) in \cite{wegg_etal_15} for the thin bar. For Model~3, the vertical scale height varies from $\sim 5^\degree$ (750 pc) in the center to about $15^\degree$ (1.2 kpc) at the bar end, which are much larger than the observational results.

Our simulations do not include gas and the subsequent star formation. This may partially account for the larger scale height values in our $N$-body simulations as compared to \cite{wegg_etal_15}. Our simulations mainly represent the old stellar population. The newly formed stars tend to be close to the disk plane \citep{debatt_etal_15}, and may gradually compress the thickness of the older population, possibly resulting in a smaller vertical scale height of the older population.

\section{DISCUSSION}
\label{sec:discussion}

\subsection{Implications of the X-shape in the Galactic Bulge}
\label{sec:X-shape}

\begin{figure}
\plotone{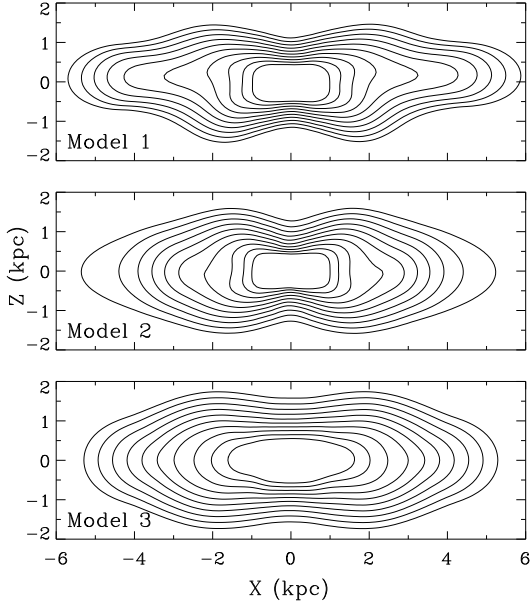}
\caption{Iso-density contours of Model 1 (top), Model 2 (middle) and Model 3 (bottom) in the X-Z plane across the bar major axis.}
\label{fig:cont_xz}
\end{figure}

Since the discovery and following numerical confirmation of the X-shaped structure in the Galactic bulge, kinematics and geometry of the X-shape have been widely investigated using both observations and simulations. Apparently the X-shape is tightly connected to the peanut component, but its three-dimensional morphological properties are still unclear. In this study, we derive the iso-density surfaces of numerical bars, aiming to differentiate between the X-shaped structures and the iso-density surfaces at various density levels. However, from our results, there is no clear morphological evidence for a component shaped as a simple letter X.

To test the possibility of the over-smoothing effect in the previous iso-density surfaces, we use a much smaller initial kernel size (0.5 kpc) to derive the spatial density. In this case, the minimum adaptive kernel size in the B/PS bulge is $\sim$ 0.2 kpc, which is small enough to reveal the X-shape. As expected, the iso-density surfaces are very noisy, but the overall shape remains peanut-like. In an extreme case, we even adopt a non-adaptive kernel size of 0.1 kpc, i.e., the kernel size remains constant. The results are consistent. We also superimpose 2000 snapshots (with the bar major axis properly aligned), but we still do not see the letter X.

This seems contradictory to our impression of a clear X-shape from the edge-on perspective of previous observations and simulations. Although the X-shaped structure seemingly exists in previous works, from the iso-density surfaces, the X-shape is not significant, which is more like a reflection of a peanut. To understand this paradoxical inconsistency, we perform a simple test in Figure~\ref{fig:cont_xz}, which shows the iso-density contours in the slice across the bar major axis in the $X-Z$ plane. It is important to keep in mind that the true 3D morphology of the X-shape should not be visualized as a simple letter X with four or eight conspicuous arms sticking out. We suspect a visually enhanced X-shape, which is probably due to the pinched concave iso-density contours in the inner region of the peanut bulge \citep{pfe_fri_91}. From Figure~\ref{fig:cont_xz}, close to the minor axis, the iso-density contours are denser than the sparse outer region, amplifying the visual illusion of the X-shape.

The true 3D shape or structure of iso-density surfaces should really be more like a peanut. Figures~\ref{fig:iso-density_out} and \ref{fig:iso-density_in} clearly show that a buckled bar is composed of three components with increasing sizes: a central boxy core, a peanut bulge and an extended thin bar. It is worth emphasizing that lines of sight from the solar perspective passing through the two humps of the peanut naturally induce a bimodal density distribution, which was interpreted as evidence for the X-shaped structure \citep{mcw_zoc_10}.

Orbital studies are important to understand the bar and B/PS morphological properties. For a three-dimensional buckled bar, the backbone orbit family is the x1 tree, i.e., the x1 family with the tree of three-dimensional orbits derived from it \citep{pfe_fri_91}. The banana-like orbits were previously suggested as the main backbone of the X-shape and B/PS bulge \citep{patsis_etal_02, skokos_etal_02, pat_kat_14}. Recent work by \cite{portai_etal_15b} classified orbital families in the peanut/X-shaped bulges and suggested brezel-like orbits, whose origin could be closely related to the x1mul2 family \citep{pat_kat_14}, as the main contributor to the X-shape. \cite{qin_etal_15} also did not find clear signatures of streaming motions along the banana orbits in the X-shaped structure of the \cite{shen_etal_10} model. Apparently, more studies on vertical resonant heating and its influence on the orbital structures (e.g., \citealt{quille_etal_14}) are needed to understand the formation of the X-shape and its orbital composition, and to connect it to the observed kinematics \citep{vasque_etal_13, zoccal_etal_14}.

\subsection{A Central Boxy Core in the Galactic Bulge?}
\label{sec:core}

In the central region ($<$ 1 kpc) of the peanut structure, the iso-density surface suggests a boxy core (like an oblong tablet), which previously has not been widely discussed in the literature. From the solar perspective, sight lines passing through the central boxy core do not display bimodal distributions. It extends to $\sim$ 500 pc from the disk plane ($|b| \sim 4^\degree$). In fact, observers do not detect the RC split (i.e., the X-shape) at low Galactic latitudes. \cite{mcw_zoc_10} found that, at $|b| \sim 5^\degree$ the two RCs density peaks seem to merge together. \cite{weg_ger_13a} also concluded that the X-shape only exists at larger latitudes above the disk plane. In fact, Figures 18 and 19 in \cite{weg_ger_13a} have already shown a morphological transition of the bar from an outer peanut structure to an inner boxy core (also see \citealt{gardne_etal_14}). \cite{qin_etal_15} also concluded that the double peaks in distance distributions towards the Galactic bulge are single-peaked within $|b| \sim 4^\degree$.

As shown in Figures 2 and 3 from \cite{portai_etal_15b}, different orbits populate different radial ranges. The edge-on projection and radial range of class A orbit might be consistent with the central boxy core identified in this work based on numerical models. However, the possible existence of such a boxy core in the Galactic bulge needs to be checked in future observations. Dust extinction and contamination from foreground disk stars at lower Galactic latitudes could potentially weaken this double peak phenomenon. The possible presence of a small classical bulge may also blur the double peak features close to the Galactic plane.

\section{SUMMARY}
\label{sec:summary}

In this work, we investigated the spatial morphology of the X-shaped structure and the peanut bulge based on three bar models with different amplitudes of the buckling instability. Based on the iso-density surfaces, the morphology of the buckled bar transitions at different density levels, from a central boxy core (tablet-like), to a vertically thickened B/PS bulge, and then to an extended thin bar.

The 3D iso-density surfaces do not reveal a simple letter X. The real 3D structure of the X-shape should be more peanut-like. The visual perception of an ``X'' is probably enhanced by the pinched concave isophotes. Note that the peanut bulge itself can qualitatively reproduce the observed bimodal distributions which were used to infer the existence of the X-shape in the Galactic bulge.

The central boxy core extends to almost 500 pc away from the Galactic plane ($|b| \leq 4^\degree$). Sight lines through this component do not produce bimodal distributions. This is consistent with observations, but the existence of this component needs to be confirmed by future observations.


We thank the anonymous referee for the constructive suggestions that help improve the paper. We also thank Jerry Sellwood and Min Du for their helpful discussions. The research presented here is partially supported by the 973 Program of China under grant no. 2014CB845700, by the National Natural Science Foundation of China under grant nos.11333003, 11322326, 11403072, and by the Strategic Priority Research Program ``The Emergence of Cosmological Structures'' (no. XDB09000000) of the Chinese Academy of Sciences. ZYL is sponsored by Shanghai Yangfan Research Grant (no. 14YF1407700). This work made use of the facilities of the Center for High Performance Computing at Shanghai Astronomical Observatory.


\begin{thebibliography}{}

\bibitem[{{Athanassoula}(2005)}]{athana_05}
{Athanassoula}, E. 2005, \mnras, 358, 1477

\bibitem[{{Benjamin} {et~al.}(2005){Benjamin}, {Churchwell}, {Babler},
  {Indebetouw}, {Meade}, {Whitney}, {Watson}, {Wolfire}, {Wolff}, {Ignace},
  {Bania}, {Bracker}, {Clemens}, {Chomiuk}, {Cohen}, {Dickey}, {Jackson},
  {Kobulnicky}, {Mercer}, {Mathis}, {Stolovy}, \& {Uzpen}}]{benjam_etal_05}
{Benjamin}, R.~A., {Churchwell}, E., {Babler}, B.~L., {et~al.} 2005, \apjl,
  630, L149

\bibitem[{{Bissantz} \& {Gerhard}(2002)}]{bis_ger_02}
{Bissantz}, N., \& {Gerhard}, O. 2002, \mnras, 330, 591

\bibitem[{{Blitz} \& {Spergel}(1991)}]{bli_spe_91}
{Blitz}, L., \& {Spergel}, D.~N. 1991, \apj, 379, 631

\bibitem[{{Bureau} \& {Freeman}(1999)}]{bur_fre_99}
{Bureau}, M., \& {Freeman}, K.~C. 1999, \aj, 118, 126

\bibitem[{{Cao} {et~al.}(2013){Cao}, {Mao}, {Nataf}, {Rattenbury}, \&
  {Gould}}]{cao_etal_13}
{Cao}, L., {Mao}, S., {Nataf}, D., {Rattenbury}, N.~J., \& {Gould}, A. 2013,
  \mnras, 434, 595

\bibitem[{{Combes} {et~al.}(1990){Combes}, {Debbasch}, {Friedli}, \&
  {Pfenniger}}]{combes_etal_90}
{Combes}, F., {Debbasch}, F., {Friedli}, D., \& {Pfenniger}, D. 1990, \aap,
  233, 82

\bibitem[{{Combes} \& {Sanders}(1981)}]{com_san_81}
{Combes}, F., \& {Sanders}, R.~H. 1981, \aap, 96, 164

\bibitem[{{Debattista} {et~al.}(2005){Debattista}, {Carollo}, {Mayer}, \&
  {Moore}}]{debatt_etal_05}
{Debattista}, V.~P., {Carollo}, C.~M., {Mayer}, L., \& {Moore}, B. 2005, \apj,
  628, 678

\bibitem[{{Debattista} {et~al.}(2006){Debattista}, {Mayer}, {Carollo}, {Moore},
  {Wadsley}, \& {Quinn}}]{debatt_etal_06}
{Debattista}, V.~P., {Mayer}, L., {Carollo}, C.~M., {et~al.} 2006, \apj, 645,
  209

\bibitem[{{Debattista} {et~al.}(2015){Debattista}, {Ness}, {Earp}, \&
  {Cole}}]{debatt_etal_15}
{Debattista}, V.~P., {Ness}, M., {Earp}, S.~W.~F., \& {Cole}, D.~R. 2015,
  \apjl, 812, L16

\bibitem[{{Erwin} \& {Debattista}(2013)}]{erw_deb_13}
{Erwin}, P., \& {Debattista}, V.~P. 2013, \mnras, 431, 3060

\bibitem[{{Gardner} {et~al.}(2014){Gardner}, {Debattista}, {Robin},
  {V{\'a}squez}, \& {Zoccali}}]{gardne_etal_14}
{Gardner}, E., {Debattista}, V.~P., {Robin}, A.~C., {V{\'a}squez}, S., \&
  {Zoccali}, M. 2014, \mnras, 438, 3275

\bibitem[{{Kormendy} \& {Kennicutt}(2004)}]{kor_ken_04}
{Kormendy}, J., \& {Kennicutt}, Jr., R.~C. 2004, \araa, 42, 603

\bibitem[{{Kuijken} \& {Merrifield}(1995)}]{kui_mer_95}
{Kuijken}, K., \& {Merrifield}, M.~R. 1995, \apjl, 443, L13

\bibitem[{{Laurikainen} {et~al.}(2014){Laurikainen}, {Salo}, {Athanassoula},
  {Bosma}, \& {Herrera-Endoqui}}]{laurik_etal_14}
{Laurikainen}, E., {Salo}, H., {Athanassoula}, E., {Bosma}, A., \&
  {Herrera-Endoqui}, M. 2014, \mnras, 444, L80

\bibitem[{{Li} \& {Shen}(2012)}]{li_she_12}
{Li}, Z.-Y., \& {Shen}, J. 2012, \apjl, 757, L7

\bibitem[{{Li} {et~al.}(2014){Li}, {Shen}, {Rich}, {Kunder}, \&
  {Mao}}]{li_etal_14}
{Li}, Z.-Y., {Shen}, J., {Rich}, R.~M., {Kunder}, A., \& {Mao}, S. 2014, \apjl,
  785, L17

\bibitem[{{L{\'o}pez-Corredoira} {et~al.}(2007){L{\'o}pez-Corredoira},
  {Cabrera-Lavers}, {Mahoney}, {Hammersley}, {Garz{\'o}n}, \&
  {Gonz{\'a}lez-Fern{\'a}ndez}}]{lopezc_etal_07}
{L{\'o}pez-Corredoira}, M., {Cabrera-Lavers}, A., {Mahoney}, T.~J., {et~al.}
  2007, \aj, 133, 154

\bibitem[{{Martinez-Valpuesta} \& {Gerhard}(2011)}]{mar_ger_11}
{Martinez-Valpuesta}, I., \& {Gerhard}, O. 2011, \apjl, 734, L20

\bibitem[{{Martinez-Valpuesta} {et~al.}(2006){Martinez-Valpuesta}, {Shlosman},
  \& {Heller}}]{martin_etal_06}
{Martinez-Valpuesta}, I., {Shlosman}, I., \& {Heller}, C. 2006, \apj, 637, 214

\bibitem[{{McWilliam} \& {Zoccali}(2010)}]{mcw_zoc_10}
{McWilliam}, A., \& {Zoccali}, M. 2010, \apj, 724, 1491

\bibitem[{{Molloy} {et~al.}(2015{\natexlab{a}}){Molloy}, {Smith}, {Evans}, \&
  {Shen}}]{molloy_etal_15b}
{Molloy}, M., {Smith}, M.~C., {Evans}, N.~W., \& {Shen}, J. 2015{\natexlab{a}},
  ArXiv e-prints, arXiv:1505.04245

\bibitem[{{Molloy} {et~al.}(2015{\natexlab{b}}){Molloy}, {Smith}, {Shen}, \&
  {Wyn Evans}}]{molloy_etal_15a}
{Molloy}, M., {Smith}, M.~C., {Shen}, J., \& {Wyn Evans}, N.
  2015{\natexlab{b}}, \apj, 804, 80

\bibitem[{{Nataf} {et~al.}(2010){Nataf}, {Udalski}, {Gould}, {Fouqu{\'e}}, \&
  {Stanek}}]{nataf_etal_10}
{Nataf}, D.~M., {Udalski}, A., {Gould}, A., {Fouqu{\'e}}, P., \& {Stanek},
  K.~Z. 2010, \apjl, 721, L28

\bibitem[{{Nataf} {et~al.}(2015){Nataf}, {Udalski}, {Skowron}, {Szyma{\'n}ski},
  {Kubiak}, {Pietrzy{\'n}ski}, {Soszy{\'n}ski}, {Ulaczyk}, {Wyrzykowski},
  {Poleski}, {Athanassoula}, {Ness}, {Shen}, \& {Li}}]{nataf_etal_15}
{Nataf}, D.~M., {Udalski}, A., {Skowron}, J., {et~al.} 2015, \mnras, 447, 1535

\bibitem[{{Ness} {et~al.}(2012){Ness}, {Freeman}, {Athanassoula},
  {Wylie-De-Boer}, {Bland-Hawthorn}, {Lewis}, {Yong}, {Asplund}, {Lane},
  {Kiss}, \& {Ibata}}]{ness_etal_12}
{Ness}, M., {Freeman}, K., {Athanassoula}, E., {et~al.} 2012, \apj, 756, 22

\bibitem[{{Patsis} \& {Katsanikas}(2014)}]{pat_kat_14}
{Patsis}, P.~A., \& {Katsanikas}, M. 2014, \mnras, 445, 3525

\bibitem[{{Patsis} {et~al.}(2002){Patsis}, {Skokos}, \&
  {Athanassoula}}]{patsis_etal_02}
{Patsis}, P.~A., {Skokos}, C., \& {Athanassoula}, E. 2002, \mnras, 337, 578

\bibitem[{{Pfenniger} \& {Friedli}(1991)}]{pfe_fri_91}
{Pfenniger}, D., \& {Friedli}, D. 1991, \aap, 252, 75

\bibitem[{{Pietrukowicz} {et~al.}(2014){Pietrukowicz}, {Kozlowski}, {Skowron},
  {Soszynski}, {Udalski}, {Poleski}, {Wyrzykowski}, {Szymanski}, {Pietrzynski},
  {Ulaczyk}, {Mroz}, {Skowron}, \& {Kubiak}}]{pietru_etal_14}
{Pietrukowicz}, P., {Kozlowski}, S., {Skowron}, J., {et~al.} 2014, ArXiv
  e-prints, arXiv:1412.4121

\bibitem[{{Portail} {et~al.}(2015{\natexlab{a}}){Portail}, {Wegg}, \&
  {Gerhard}}]{portai_etal_15b}
{Portail}, M., {Wegg}, C., \& {Gerhard}, O. 2015{\natexlab{a}}, ArXiv e-prints,
  arXiv:1503.07203

\bibitem[{{Portail} {et~al.}(2015{\natexlab{b}}){Portail}, {Wegg}, {Gerhard},
  \& {Martinez-Valpuesta}}]{portai_etal_15a}
{Portail}, M., {Wegg}, C., {Gerhard}, O., \& {Martinez-Valpuesta}, I.
  2015{\natexlab{b}}, \mnras, 448, 713

\bibitem[{{Qin} {et~al.}(2015){Qin}, {Shen}, {Li}, {Mao}, {Smith}, {Rich},
  {Kunder}, \& {Liu}}]{qin_etal_15}
{Qin}, Y., {Shen}, J., {Li}, Z.-Y., {et~al.} 2015, \apj, 808, 75

\bibitem[{{Quillen} {et~al.}(2014){Quillen}, {Minchev}, {Sharma}, {Qin}, \& {Di
  Matteo}}]{quille_etal_14}
{Quillen}, A.~C., {Minchev}, I., {Sharma}, S., {Qin}, Y.-J., \& {Di Matteo}, P.
  2014, \mnras, 437, 1284

\bibitem[{{Raha} {et~al.}(1991){Raha}, {Sellwood}, {James}, \&
  {Kahn}}]{raha_etal_91}
{Raha}, N., {Sellwood}, J.~A., {James}, R.~A., \& {Kahn}, F.~D. 1991, \nat,
  352, 411

\bibitem[{{Rattenbury} {et~al.}(2007){Rattenbury}, {Mao}, {Sumi}, \&
  {Smith}}]{ratten_etal_07}
{Rattenbury}, N.~J., {Mao}, S., {Sumi}, T., \& {Smith}, M.~C. 2007, \mnras,
  378, 1064

\bibitem[{{Saito} {et~al.}(2011){Saito}, {Zoccali}, {McWilliam}, {Minniti},
  {Gonzalez}, \& {Hill}}]{saito_etal_11}
{Saito}, R.~K., {Zoccali}, M., {McWilliam}, A., {et~al.} 2011, \aj, 142, 76

\bibitem[{{Sellwood}(2014)}]{sellwo_14}
{Sellwood}, J.~A. 2014, Reviews of Modern Physics, 86, 1

\bibitem[{{Sellwood} \& {McGaugh}(2005)}]{sel_mcg_05}
{Sellwood}, J.~A., \& {McGaugh}, S.~S. 2005, \apj, 634, 70

\bibitem[{{Sellwood} \& {Merritt}(1994)}]{sel_mer_94}
{Sellwood}, J.~A., \& {Merritt}, D. 1994, \apj, 425, 530

\bibitem[{{Shen} \& {Li}(2015)}]{she_li_15}
{Shen}, J., \& {Li}, Z.-Y. 2015, ArXiv e-prints, arXiv:1504.05136

\bibitem[{{Shen} {et~al.}(2010){Shen}, {Rich}, {Kormendy}, {Howard}, {De
  Propris}, \& {Kunder}}]{shen_etal_10}
{Shen}, J., {Rich}, R.~M., {Kormendy}, J., {et~al.} 2010, \apjl, 720, L72

\bibitem[{{Shen} \& {Sellwood}(2004)}]{she_sel_04}
{Shen}, J., \& {Sellwood}, J.~A. 2004, \apj, 604, 614

\bibitem[{{Silverman}(1986)}]{silver_86}
{Silverman}, B.~W. 1986, {Density estimation for statistics and data analysis}

\bibitem[{{Skokos} {et~al.}(2002){Skokos}, {Patsis}, \&
  {Athanassoula}}]{skokos_etal_02}
{Skokos}, C., {Patsis}, P.~A., \& {Athanassoula}, E. 2002, \mnras, 333, 847

\bibitem[{{V{\'a}squez} {et~al.}(2013){V{\'a}squez}, {Zoccali}, {Hill},
  {Renzini}, {Gonz{\'a}lez}, {Gardner}, {Debattista}, {Robin}, {Rejkuba},
  {Baffico}, {Monelli}, {Motta}, \& {Minniti}}]{vasque_etal_13}
{V{\'a}squez}, S., {Zoccali}, M., {Hill}, V., {et~al.} 2013, \aap, 555, A91

\bibitem[{{Wegg} \& {Gerhard}(2013)}]{weg_ger_13a}
{Wegg}, C., \& {Gerhard}, O. 2013, \mnras, 435, 1874

\bibitem[{{Wegg} {et~al.}(2015){Wegg}, {Gerhard}, \& {Portail}}]{wegg_etal_15}
{Wegg}, C., {Gerhard}, O., \& {Portail}, M. 2015, \mnras, 450, 4050

\bibitem[{{Zoccali} {et~al.}(2014){Zoccali}, {Gonzalez}, {Vasquez}, {Hill},
  {Rejkuba}, {Valenti}, {Renzini}, {Rojas-Arriagada}, {Martinez-Valpuesta},
  {Babusiaux}, {Brown}, {Minniti}, \& {McWilliam}}]{zoccal_etal_14}
{Zoccali}, M., {Gonzalez}, O.~A., {Vasquez}, S., {et~al.} 2014, \aap, 562, A66

\end{thebibliography}

\end{document}